\documentclass[aps,prl,twocolumn,superscriptaddress,footinbib]{revtex4-1}
\usepackage{bm}
\usepackage{mathrsfs}
\usepackage{amsmath}
\usepackage{amssymb}
\usepackage{graphicx}
\usepackage{amsfonts}
\usepackage{amsthm}
\usepackage{color}
\usepackage{dcolumn}
\usepackage{txfonts}
\usepackage{times}
\usepackage[normalem]{ulem}

\begin{document}

\title{Dynamical Scaling Laws of Out-of-Time-Ordered Correlators}
\author{Bo-Bo Wei}
\email{Corresponding author: weibobo@cuhk.edu.cn}
\affiliation{School of Science and Engineering, The Chinese University of Hong Kong, Shenzhen, Shenzhen 518172, China}
\affiliation{Center for Quantum Computing, Peng Cheng Laboratory, Shenzhen 518055, China}
\author{Gaoyong Sun}
\email{Corresponding author: gysun@nuaa.edu.cn}
\affiliation{College of Science, Nanjing University of Aeronautics and Astronautics, Nanjing 211106, China}
\author{Myung-Joong Hwang}
\email{Corresponding author: myungjoong.hwang@uni-ulm.de}
\affiliation{Institut f\"{u}r Theoretische Physik and IQST, Albert-Einstein-Allee 11, Universit\"{a}t Ulm, D-89069 Ulm, Germany}

\begin{abstract}
The out-of-time-ordered correlator (OTOC) is central to the understanding of information scrambling in quantum many-body systems. In this work, we show that the OTOC in a quantum many-body system close to its critical point obeys dynamical scaling laws which are specified by a few universal critical exponents of the quantum critical point. Such scaling laws of the OTOC imply a universal form for the butterfly velocity of a chaotic system in the quantum critical region and allow one to locate the quantum critical point and extract all universal critical exponents of the quantum phase transitions. We numerically confirm the universality of the butterfly velocity in a chaotic model, namely the transverse axial next-nearest-neighbor Ising model, and show the feasibility of extracting the critical properties of quantum phase transitions from OTOC using the Lipkin-Meshkov-Glick (LMG) model.
\end{abstract}

\maketitle

\emph{Introduction.---}  The out-of-time-ordered correlator (OTOC)~\cite{Larkin1969,Kitaev2014,Shenker2014a,Shenker2014b,Hartnoll2015,Shenker2015,Kitaev2015} has been recently proposed to characterise the way in which a local information of a quantum many-body system disperses throughout the system, a process called information scrambling~\cite{Chen2016,Banerjee2017,He2017,Shen2017,Slagle2017,Fan2017,Chen2018,Huang2018,Iyoda2018,Lin2018,Pappalardi2018,Halpern2018a,Syzranov2018,Zhang2018,Swingle2018np,McGinley2019,Alonso2019}
and also to describe chaotic behavior such as butterfly effect in quantum many-body dynamics~\cite{Roberts2015a,Roberts2015b,Maldacena2016,Aleiner2016,Blake2016b,Lucas2016,Roberts2016,Hosur2016,Gu2017,Roberts2017,Schedev2017,Yoshida2017}. For two local operators $W$ and $V$ of a quantum many-body system with Hamiltonian $H$, the OTOC is defined as~\cite{Kitaev2014,Shenker2014a}
\begin{eqnarray}\label{OTOC0}
F(t)=\langle W(t)^{\dagger}V(0)^{\dagger}W(t)V(0)\rangle,
\end{eqnarray}
where $W(t)=e^{itH}We^{-itH}$ and $V(0)$ are Heisenberg operators at time $t$ and $0$ respectively and the angular bracket denotes the expectation value over a pure state or over a thermal Gibbs state at inverse temperature $\beta=1/k_BT$ with $k_B$ being the Boltzmann constant and $T$ being the temperature. The OTOC is closely related to the squared commutator
$C(t)=\langle \left|[W(t),V(0)]\right|^2\rangle=2[1-\Re F(t)]$ for local unitary operators~\cite{Kitaev2014,Shenker2014a}. Physically, the decay of the OTOC $F(t)$  or the growth of $C(t)$ characterizes scrambling of quantum information~\cite{Sekino2008,Lashkari2013}. The OTOC also describes how a local perturbation $W$ spreads to affect the measurement of $V$ at a distance $r$, which may be viewed as a quantum butterfly effect~\cite{Shenker2014a}. Recently, several schemes~\cite{Swingle2016,Zhu2016,Yao2016,Tsuji2017,Halpern2017,Campisi2017,Swingle2018,Halpern2018b} have been proposed to measure the OTOC experimentally in a variety of physical systems~\cite{Li2017,Garttner2017,Meier2017,Landsman2018,Niknam2018}.

In equilibrium, a quantum phase transition (QPT) accompanies a long-range quantum correlation at a critical point~\cite{Sachdev2011,Cardy1996}. Understanding how such criticality and correlation in equilibrium affects the scrambling of information and the butterfly effect is an important theoretical challenge. On the one hand, in the context of holographic models dual to classical gravity, it was recently conjectured that the butterfly velocity is maximum at quantum critical point (QCP)~\cite{Shen2017}; however, whether this conjecture holds universally true for such models are yet to be understood~\cite{Blake2016a,Ling2017a,Ling2017b,Baggioli2018}. On the other hand, a long-time average of OTOC shows distinctive features between a normal and broken-symmetry phase in Ising-type spin models~\cite{Heyl2018} and it has been proposed that the long-time average (or the long-time saturated value~\cite{Duan2019}) of OTOC can be used to detect QPTs~\cite{Heyl2018,Sun2018}. The latter approach is however limited to detect the presence of broken-symmetry phase and does not allow one to extract a broad phenomenology of QPTs such as universality and scaling as well as its connection to butterfly effect from OTOC; moreover, as it relies on asymptotic dynamics on a long-time scale, a long coherence time is required in experiments, which is in general a challenging requirement for quantum many-body systems.

In this Letter, we develop dynamical scaling laws of the OTOC of both local and global operators of a strongly interacting many-body system undergoing a second-order QPT.  From such scaling laws of the OTOC, we find that the information scrambling and the butterfly velocity at the critical point are completely governed by the equilibrium critical properties of the system, namely, critical exponents and thereby the universality class. We predict a universal function for the butterfly velocity in a generic quantum chaotic and critical model and confirm this prediction using the transverse axial next-nearest-neighbor Ising models (ANNNI) model, which we numerically solve using the time-dependent density matrix renormalization group (t-DMRG) approach. Moreover, we propose a strategy that is capable of extracting critical properties of QPTs using OTOC, going beyond of a simple detection of the existence of QPTs. We note that our strategy relies only on the short, transient dynamics of OTOC rather than its asymptotic dynamics, which imposes a less stringent condition on the coherence time. To this end, we develop a finite-size scaling analysis of OTOC of global operators and test these ideas with a fully-connected Ising model known as Lipkin-Meshkov-Glick (LMG) model.

\begin{figure}
\begin{center}
\includegraphics[scale=0.038]{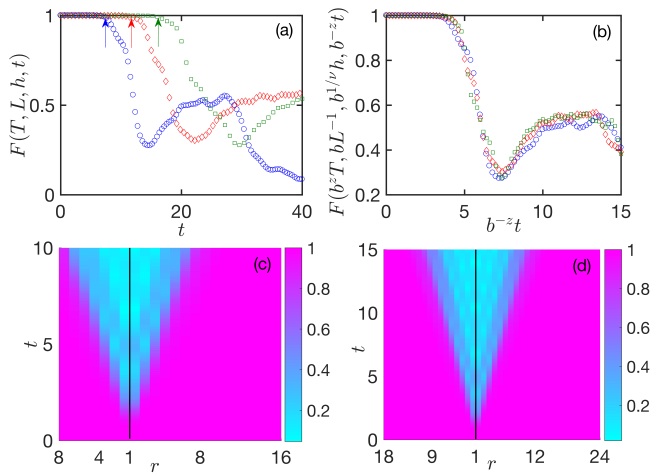}
\end{center}
\caption{(color online). Universal dynamical scaling of OTOC in the ANNNI model. (a) The OTOC $F(T,L,h,t)$ for $W=\sigma_j^x,V=\sigma_{j+L/2}^x$ with $T=0.1$ and $h=0.01$ as a function of time for different system sizes, $L=16$ (blue circle), $L=24$ (red diamond), $L=32$ (green square). The arrows mark the scrambling time. (b) Scaling invariance of OTOC given in Eq.\eqref{OTOCscalinga}. We take $L=32,T=0.1,h=0.01$ and $\nu=1$ and $z=1$. All the data for $b=1$, $4/3$, and $2$ collapse onto a single curve. (c) Density plot of the OTOC for $W=\sigma_j^x,V=\sigma_{j+r}^x$ at the critical line ($h=0$) as a function of $t$ and $r$. The left of the black solid line is for $T=0.2,L=16$ and the right is for $T=0.1,L=32$. (d) Density plot of the OTOC for $W=\sigma_j^x,V=\sigma_{j+r}^x$ at QCP ($T=h=0$) as a function of $t$ and $r$. The left of the black solid line is for $L=36$ and the right is for $L=48$. }
\label{fig:epsart1}
\end{figure}

\begin{figure}
\begin{center}
\includegraphics[scale=0.3]{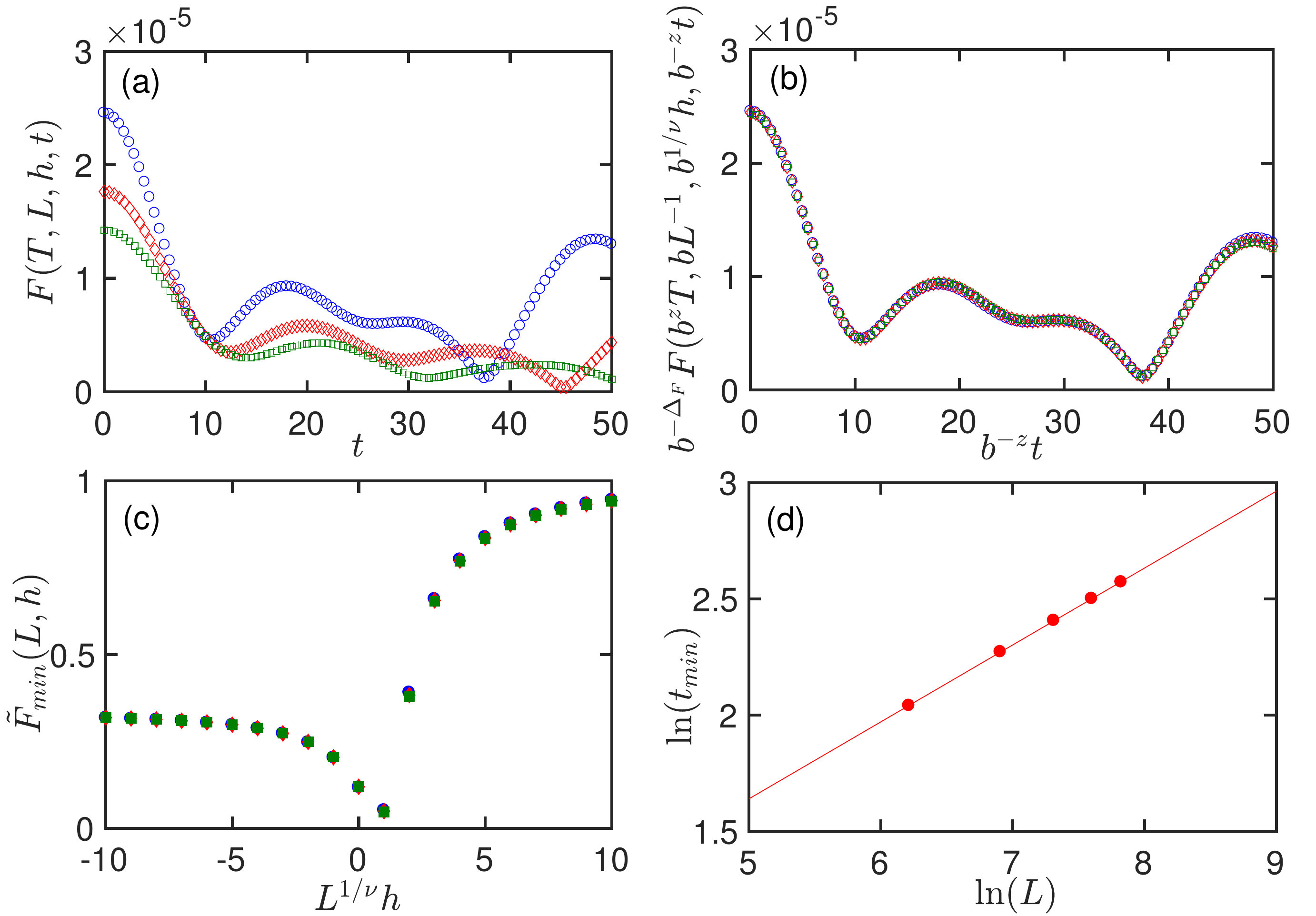}
\end{center}
\caption{(color online). Universal dynamical scaling of OTOC $F(T,L,h,t)$ for $W=V=\sigma_x=\sum_j\sigma_j^x/L$ in the LMG model at $\gamma=0.5$. (a) The OTOC with $T=0.1$ and $h=0.005$ as a function of time for different system sizes, $L=1000$ (blue circle), $L=1500$ (red diamond), $L=2000$ (green square). (b) Scaling invariance of OTOC given in Eq.~\eqref{OTOCscalingc}. We take $L=2000,T=0.02,h=0.005$, $\nu=3/2$, $z=1/3$ and $\Delta_F=4/3$. All the data for $b=1$, $4/3$, and $2$ collapse onto a single curve. (c) Data collapse of the first minimum of normalized OTOC at zero temperature as a function of $L^{1/\nu}h$ for different system sizes $N=1000,1500,2000$, where $\nu=3/2$. (d). $\ln t_{\text{min}}$ as a function of $\ln L$ at QCP. Linear fit shows that $z=0.330\pm0.001$, which agrees well with the theoretical value $z=1/3$.}
\label{fig:epsart2}
\end{figure}

\emph{Dynamical scaling of OTOC.---}
Let us consider a general quantum many-body system with Hamiltonian $H(\lambda)$ that undergoes a second order QPT at $\lambda=\lambda_c$. As the system becomes scale invariant at the critical point~\cite{Sachdev2011,Cardy1996}, we introduce scaling dimensions of $X=W, W^{\dagger}, V, V^\dagger$, operators defining the OTOC in Eq.~\eqref{OTOC0} as $[X]=\Delta_X$. Namely, under a scaling transformation, $r\rightarrow r'=r/b$, the operator $X$ transforms as $X\rightarrow b^{-\Delta_X}X$~\cite{Sachdev2011,Cardy1996}. We note that the scale transformation with $b>1$ corresponds to grouping $b$ lattice sites into a single block. Relevant parameters determining the OTOC near a critical point are the temperature of the initial thermal state $T$, the inverse of the system length $L^{-1}$, the distance of the control parameter to the quantum critical point $h=|\lambda-\lambda_c|$, the time $t$ and the distance $r$ between local operators $W$ and $V$, that is, $F=F(T,L^{-1},h,r,t)$. The scaling dimensions of these variables $T$, $L$, $h$, $t$, and $r$ are well-known to be $[T]=z$, $[L^{-1}]=1$, $[h]=1/\nu$, $[t]=-z$, and $[r]=-1$ where $\nu$ is the correlation length critical exponent, $\xi\sim h^{-\nu}$ with $\xi$ being the correlation length and $z$ is the dynamical critical exponent, $\tau\sim\xi^z$ with $\tau$ being the correlation time. By applying a scaling transformation based on scaling dimensions of operators and variables introduced above, we have
\begin{eqnarray}
F(T,L^{-1},h,r,t)&=&b^{-\Delta_F} \Psi\left(b^zT,b L^{-1},b^{1/\nu}h,b^{-1}r,b^{-z}t\right)\label{OTOCscaling}
\end{eqnarray}
for a critical system in a scaling limit ($L^{-1}, h\ll1$). Here $\Delta_F=\Delta_W+\Delta_{W^{\dagger}}+\Delta_V+\Delta_{V^{\dagger}}$ is the scaling exponent for the OTOC and $\Psi$ is a universal scaling function. Now we make several remarks about Eq.~\eqref{OTOCscaling}:\\
\textbf{(I)}.~For local unitary operators $W$ and $V$, we have $W^{\dagger}W=1$ and $ V^{\dagger}V=1$, both of which are invariant under scaling transformation and we then have $F(t=0)=1$ and Eq.~\eqref{OTOCscaling} simplifies to
\begin{eqnarray}
F(T,L^{-1},h,r,t)&=&\Phi\left(b^zT,b L^{-1},b^{1/\nu}h,b^{-1}r,b^{-z}t\right)\label{OTOCscalinga}.
\end{eqnarray}
This suggests that the OTOC for local unitary operators close to QCP is invariant under scaling transformations.
\\\textbf{(II)}.~If $W$ and $V$ are global operators, such as in Ref.~\cite{Heyl2018,Duan2019}, the OTOC does not depend on distance and Eq.~\eqref{OTOCscaling} reduces into
\begin{eqnarray}
F(T,L^{-1},h,t)&=&b^{-\Delta_F} \Psi\left(b^zT,b L^{-1},b^{1/\nu}h,b^{-z}t\right)\label{OTOCscalingc}.
\end{eqnarray}
Below, we numerically confirm the validity of the predicted dynamical scaling of OTOC, Eq.~\eqref{OTOCscalinga} and \eqref{OTOCscalingc}, using a non-integrable, critical Ising-type model and an Ising model with infinite-range interactions. [See Fig.~\ref{fig:epsart1} (a, b) and Fig.~\ref{fig:epsart2} (a, b)], which are the first results of this letter and provide a basis to investigate further the critical properties of OTOC.

\emph{Universality of butterfly effect in chaotic, critical quantum systems.---} For a quantum chaotic system, the OTOC $F(T,L^{-1},h,r,t)$  for local operators separated by a distance $r$ typically exhibits a ballistic traveling and the scrambling time $t_s$ linearly depends on the distance $r$~\cite{Sekino2008,Swingle2018np}. The butterfly velocity defined as $v_B=r/t_s$ quantifies the spreading of information in quantum systems; for example, we have $F(r,t)=1$ for $r\gg v_Bt$ and $F(r,t)=0$ for $r\ll v_Bt$. The aim of this section is to investigate the critical properties of butterfly velocity and to demonstrate its universality.

In the quantum critical region where the temperature is the dominant scale~\cite{Sachdev2011}, we set $b^{z}T=1$ in Eq.~\eqref{OTOCscalinga} and obtain the dynamical scaling form
\begin{eqnarray}\label{T1}
F(T,L^{-1},h,r,t)&=&\Phi_1\left(hT^{-1/(\nu z)},T^{-1/z} L^{-1},T^{1/z}r,Tt\right).
\end{eqnarray}
Here $\Phi_1$ is a universal scaling function for the OTOC onto which all data for $F(T,L^{-1},h,r,t)$ collapse when properly rescaled as suggested by Eq.~\eqref{T1}. To uncover the critical properties of butterfly velocity in Eq.~\eqref{T1}, we consider OTOCs for two different sets of parameters $(T_1,L_1,h_1,r_1)$ and $(T_2,L_2,h_2,r_2)$. According to Eq.~\eqref{T1}, if these two parameter sets are chosen such that the arguments of scaling functions are identical, i.e., $h_1T_1^{-1/(\nu z)}=h_2T_2^{-1/(\nu z)}$, $T_1^{-1/z} L_1^{-1}=T_2^{-1/z} L_2^{-1}$, and $T_1^{1/z}r_1=T_2^{1/z}r_2$, then the renormalized scrambling time $Tt_s$  should also be identical $T_1t_{s1}=T_2t_{s2}$. By dividing the last two equalities, we get
\begin{equation}
 v_B(T_1,h_1,L_1)T_1^{1/z-1}=v_B(T_2,h_2,L_2)T_2^{1/z-1}
\end{equation}
In order to satisfy the above condition, the butterfly velocity should contain a factor $T^{1-1/z}$ in addition to a function that depends only on the rescaled variables of $hT^{-1/(\nu z)}$ and $T^{-1/z} L^{-1}$; namely, the butterfly velocity of a chaotic quantum many-body system in the quantum critical region should take the form
\begin{eqnarray}\label{velocity}
v_B(T,h,L)=\mathcal{G}\left(hT^{-1/(\nu z)},T^{-1/z} L^{-1}\right)T^{1-1/z},
\end{eqnarray}
where $\mathcal{G}(x,y)$ is a universal scaling function.

From Eq.~\eqref{velocity}, we can obtain the finite-size scaling exponent of OTOC along the quantum critical trajectory (line with $h=0$ in the quantum critical region). For $h=0$, we have
\begin{eqnarray}\label{velocity2}
v_B(T,0,L)=\mathcal{Q}\left(T^{-1/z} L^{-1}\right)T^{1-1/z}.
\end{eqnarray}
Here $\mathcal{Q}(y)\equiv\mathcal{G}(0,y)$. In order to eliminate temperature dependence when $T\rightarrow0$ and $L<\infty$, it is necessary to have  $\lim_{y\rightarrow\infty}\mathcal{Q}(y)=y^{\gamma}$ with $\gamma=z-1$. Thus the butterfly velocity at the quantum critical point ($h=0,T=0$) for a finite system size is
\begin{eqnarray}\label{velocity2a}
v_B(0,0,L)\propto L^{-(z-1)}.
\end{eqnarray}
Eq.~\eqref{velocity} to Eq.~\eqref{velocity2a} are the predictions for the butterfly velocity in the quantum critical region from the scaling invariance of the OTOC for local unitary operators. Note that if the dynamical critical exponent $z<1$, then the butterfly velocity in Eq.~\eqref{velocity} to Eq.~\eqref{velocity2a} diverge at quantum critical point or in the thermodynamic limit. But, the causality requires that $v_B$ should be bounded. Thus Eq.~\eqref{velocity} to Eq.~\eqref{velocity2a} are valid only for $z\geq1$.

In the thermodynamic limit, $L\rightarrow\infty$, we have
\begin{eqnarray}\label{velocity1}
v_{B,\infty}(T,h)=\mathcal{K}\left(hT^{-1/(\nu z)}\right)T^{1-1/z}.
\end{eqnarray}
Here $\mathcal{K}(x)\equiv\mathcal{G}(x,0)$. To eliminate temperature dependence when $h\neq0$ and $T\rightarrow0$, it is necessary to have $\lim_{x\rightarrow\infty}\mathcal{K}(x)=x^{\alpha}$ with $\alpha=\nu(z-1)$. Thus the butterfly velocity along the quantum critical trajectory ($h=0$) and at zero temperature are respectively
\begin{eqnarray}
v_{B,\infty}(T,0)&\propto& T^{1-1/z},\label{velocity1a}\\
v_{B,\infty}(0,h)&\propto& h^{\nu(z-1)}.\label{velocity1b}
\end{eqnarray}
From Eq.~\eqref{velocity1a}, we predict that for a chaotic quantum many-body system of Ising universality class with $z=1$~\cite{Sachdev2011} we have $v_B\propto T^0$; that is, the butterfly velocity becomes temperature independent. On the other hand, for superfluid transition of Bose-Hubbard model~\cite{Sachdev2011}, $z=2$, we predict that the butterfly velocity increases as a square-root of temperature, $v_B\propto \sqrt{T}$. Finally, Eq.~\eqref{velocity1b} shows that the butterfly velocity vanishes right at QCP ($h=T=0$) and increases as one deviates from the QCP at the zero temperature. It is interesting to note that Eq.~\eqref{velocity1} and Eq.~\eqref{velocity1a} agree with previous predictions for a class of holographic models with classical gravity near the QCP~\cite{Roberts2016,Blake2016a,Ling2017a}; meanwhile, our derivation is based on the scaling and universality of QPTs and should be valid for any chaotic many-body systems undergoing a QPT. The dynamical scaling laws of OTOC described so far is completely general and in the following we support their validity using a specific quantum chaotic many-body model.

\emph{The ferromagnetic axial next-nearest-neighbor Ising model.---}
To illustrate the universal critical properties of OTOC and of the butterfly velocity, we focus on the the transverse axial next-nearest-neighbor Ising (ANNNI) model. The model is a paradigmatic example of a quantum many-body model exhibiting both non-integrability, thus leading to a possibility of chaotic behavior, and quantum phase transition. The Hamiltonian of the ANNNI model is~\cite{Karrasch2013,Alba2017}
\begin{eqnarray}
H=-J\sum_{j=1}^L[\sigma_j^x\sigma_{j+1}^x+\Delta\sigma_j^x\sigma_{j+2}^x+\lambda\sigma_j^z].
\end{eqnarray}
Here $\sigma_j^{\alpha}$ is the Pauli matrix at site $j$ along $\alpha=x,y,z$ directions, $J$ is the ferromagnetic coupling, $\lambda$ is a transverse magnetic field, and $\Delta$ quantifies the strength of the next-nearest neighbor interaction. For $\Delta=0$, the model recovers the transverse field Ising chain. For $\Delta>-0.5$ ($\Delta\neq0$), on the other hand, the model is non-integrable and presents a QPT of Ising universality class with critical exponents $\nu=z=1$ and thus we choose $\Delta=-0.3$ and $J=1$ in all numerical simulations. We choose single-site spin operators $W=\sigma^x_i$ and $V=\sigma^x_j$ that are separated by a distance $r=j-i$ as operators defining the OTOC in Eq.~\eqref{fig:epsart1}, which are unitary operators. This choice goes beyond the case previously studied in Ref.~\cite{Heyl2018} where an identical local spin operator $W=V$ is chosen and allows one to investigate the butterfly effect at a critical point. Finally, for a numerical simulation, we utilize t-DMRG approach~\cite{Schollwoeck2011} which allows us to simulate the dynamics of a 1D spin chain of length up to 48 spins.

Figure~\ref{fig:epsart1} shows the OTOC in the ANNNI model close to QCP. Fig.~1(a) presents OTOC with $W=\sigma_j^x,V=\sigma_{j+L/2}^x$ of the thermal equilibrium state at $T=0.1$ and $h=0.01$ as a function of time for different lattice sizes $L=16,24,32$. One can see that the OTOC remains to be $1$ until the scrambling time $t_s$ at which OTOC starts to decay, as indicated by the arrows in Fig.~\ref{fig:epsart1}(a). As the distance between $W$ and $V$ increases for bigger system size, the scrambling time increases too [Fig.~\ref{fig:epsart1}(a)]. Since $W$ and $V$ are local unitary operators, the OTOC should obey scaling invariance described by Eq.~\eqref{OTOCscalinga}. We confirm that this is indeed the case by showing that all data points collapse into a single scaling function when properly rescaled [Fig.~\ref{fig:epsart1}(b)].

Next, we consider the OTOC with $M=\sigma_j^x$ and $V=\sigma_{j+r}^x$ with varying distance $r$. According to Eq.~\eqref{velocity2}, for ANNNI with $z=1$, the butterfly velocity at $h=0$ is a universal function of $TL$. To confirm this prediction, in Fig.~\ref{fig:epsart1}(c), we present the density plot of the OTOC with $M=\sigma_j^x,V=\sigma_{j+r}^x$ of the thermal equilibrium state with $h=0$ as a function of $t$ and $r$. The density plot on the left and the right side of the solid line in Fig.~\ref{fig:epsart1}(c) is for $T=0.2,L=16$ and $T=0.1,L=32$, which have an identical value for the scaling variable $TL=3.2$. First they clearly exhibits a light-cone structure, showing the linear dependence of the scrambling time $t_s$ on the distance between local operators. The slope of the light cone on both sides are almost identical with an estimated value of butterfly velocity $v_b\sim1.37$ (left) and $v_b\sim1.42$ (right). Therefore, within the numerical error, we conclude that the butterfly velocity at the critical line $h=0$ is a universal function of $TL$, which strongly supports Eq.~\eqref{velocity2} for the ANNNI model numerically. Finally, at QCP $(T=0,h=0)$, Eq.~\eqref{velocity2a} predicts that the butterfly velocity of the ANNNI should be independent of lattice sizes $L$. In Fig.~\ref{fig:epsart1}(d), we show the density plot of the OTOC with $W=\sigma_j^x,V=\sigma_{j+r}^x$ at QCP $(T=0,h=0)$ as a function of $t$ and $r$. The black solid line in the density plot separates two different lattice sizes $L=36$ (left) and $L=48$ (right). The butterfly velocities are both estimated to be $v_b\sim1.45$, which confirms the validity of Eq.~\eqref{velocity2a}.

\emph{OTOC as a probe of QPT.---} It has been recently proposed that the long-time average of OTOC~\cite{Heyl2018,Sun2018} or the long-time saturating value of OTOC~\cite{Duan2019} can be used to detect QPTs. Here, we shall show that the dynamical scaling laws for OTOC of global operators can be utilized to locate the QCP and extract the critical exponents $\nu$ and $z$ of QPTs using the \emph{transient} dynamics of OTOC.

To avoid the prefactor in Eq.~\eqref{OTOCscalingc}, we define a normalized OTOC, $\tilde{F}\equiv F(t)/F(0)$, for global operators at zero temperature, which satisfies
\begin{eqnarray}\label{normalOTOC}
\tilde{F}(L^{-1},h,t)&=&\Psi_1\left(b L^{-1},b^{1/\nu}h,b^{-z}t\right)\label{OTOCzero}.
\end{eqnarray}
In the finite-size scaling regime $b L^{-1}\sim1$, we replace $b$ with $L$ so that we have
\begin{eqnarray}\label{fss}
\tilde{F}(L^{-1},h,t)&=&\Psi_2\left(L^{1/\nu}h,L^{-z}t\right).
\end{eqnarray}
Let us now define the first minimum of the OTOC as $\tilde{F}_{\text{min}}(L,h)=\tilde{F}(L,h,t_{\text{min}})$ where $t_{\text{min}}$ is the time at which $\tilde{F}$ takes the first local minimum~\cite{footnote,Hwang2019}. From Eq.~\eqref{fss}, we have
\begin{eqnarray}\label{fm}
\tilde{F}_{\text{min}}(L,h)=\Psi_3(L^{1/\nu}h).
\end{eqnarray}
According to Eq.~\eqref{fm}, if we plot $\tilde{F}_{\text{min}}$ as a function of control parameter $h$ for different system sizes $L$, all curves will collapse onto a single function when it is plotted against $L^{1/\nu}h$. Therefore, the correlation length exponent $\nu$ can be obtained by choosing $\nu$ so that data collapse is achieved around the critical point. The dynamical critical exponent $z$ can also be obtained from studying $t_{\text{min}}$ as a function of $L$ at QCP because Eq.~\eqref{fss} shows that at QCP,
\begin{eqnarray}\label{tmin}
t_{\text{min}}\propto L^{z}.
\end{eqnarray}
Alternatively, one could also determine $\nu$ and $z$ directly from Eq.~\eqref{fss} because it says that all curves of $\tilde{F}$ as a function of $h$ and $t$ for different system sizes $L$ shall collapse into one curve if we scale $h$ by $L^{1/\nu}h$ and $t$ by $L^{-z}t$.

The scheme for probing QPTs in terms of the normalized OTOC in Eq.~\eqref{normalOTOC} does not depend on what operators $W$ and $V$ one uses in defining OTOC as long as they are relevant scaling operators. In this sense, using the normalized OTOC is more robust than that of OTOC in Eq.~\eqref{OTOCscalingc}, which does not require a prior knowledge of scaling dimension of $W$ and $V$. Therefore, we conclude that measuring OTOC as a function of time for different sizes~\cite{Li2017,Garttner2017,Meier2017,Landsman2018,Niknam2018}, one can extract the location of QCP and all critical exponents of the QPT.

\emph{The Lipkin-Meshkov-Glick (LMG) model.---} In this section, we verify the validity of scaling laws of OTOC for a global operator in Eq.~\eqref{OTOCscalingc}, Eq.~\eqref{fm}, and Eq.~\eqref{tmin} using the fully-connected Ising model often called as the LMG model and illustrate how critical properties of QPT can be extracted from OTOC. The dynamics of LMG model have been experimentally implemented in trapped ions systems~\cite{ion1,ion2,ion3}. The Hamiltonian of LMG model is~\cite{Lipkin1965},
\begin{eqnarray}
H(\lambda)&=&-\frac{J}{L}\sum_{i<j}\Big(\sigma_i^x\sigma_j^x+\gamma\sigma_i^y\sigma_j^y\Big)-\lambda\sum_{j=1}^L\sigma_j^z.
\end{eqnarray}
Here $\gamma$ is the anisotropy of the ferromagnetic coupling in the $x$ and $y$ direction, and $\lambda$ is the magnetic field along $z$ direction. For $\gamma\neq1$, the LMG model presents a QPT~\cite{Botet1982,Botet1983,Dusuel2004} from paramagnetic phase ($\lambda>1$) to ferromagnetic phase ($\lambda<1$) at QCP $\lambda_c=1$ and the correlation length critical exponent and the dynamical critical exponents are respectively $\nu=3/2$ and $z=1/3$~\cite{Botet1982,Botet1983,Dusuel2004}.

For the OTOC, we choose a global operator $W=V=\sigma_x=\sum_j\sigma_j^x/L$. Using the scaling dimension of $\sigma_x$, $\Delta_{\sigma_x}=1/3$~\cite{Botet1982,Botet1983,Dusuel2004}, one obtains the scaling dimension of OTOC, $\Delta_F=4/3$. In Figure~\ref{fig:epsart2}, we present the numerical results for the LMG model. First, Fig.~\ref{fig:epsart2}(a) shows OTOC at $T=0.1,h=0.005$ as a function of time for different lattice sizes $L=1000,1500,2000$ respectively. If the control parameters in the OTOC are scaled by the transformation described in Eq.\eqref{OTOCscalingc}, we observe that the OTOC for different lattice sizes indeed collapse into a single curve [Fig.~\ref{fig:epsart2}(b)] as predicted by Eq.~\eqref{OTOCscalingc}. In Fig.~\ref{fig:epsart2}(c), we plot the first minimum of the normalized OTOC as a function of $L^{1/\nu}h$ at zero temperature $T=0$ and confirm that $\tilde{F}_{\text{min}}(L,h)$ in Eq.~\eqref{fm} is indeed a universal function of $L^{1/\nu}h$ for $\nu=3/2$. Finally, in Fig.~2(d), we present the logarithm of $t_{\min}$ when the OTOC presents the first minimum as a function of $\ln L$ and a linear fit of the data shows that $z=0.330\pm0.001$, which agrees to the exact solution very well~\cite{Botet1982,Botet1983,Dusuel2004}. These numerical results demonstrate that the critical exponents $z$ and $\nu$ can be extracted using the scaling laws of OTOC.

\emph{Summary and Dicussions.---}
In summary, we have shown that the OTOC in quantum many-body systems close to its critical point obeys dynamical scaling laws which are specified by a few universal critical exponents of the quantum critical point. The scaling laws of the OTOC imply a universal form for the butterfly velocity of a chaotic system in the quantum critical region and allow one to determine the quantum critical point and all universal critical exponents of quantum phase transitions. We support our results numerically in two paradigmatic quantum many-body models. Because the OTOC has been experimentally observed in a variety of physical systems \cite{Li2017,Garttner2017,Meier2017,Landsman2018,Niknam2018}, the scaling laws report in this letter could be experimentally tested and verified in the near future.

\begin{acknowledgements}
B.~B.~W.~was supported by the National Natural Science Foundation of China (Grant Number 11604220) and the President's Fund of The Chinese University of Hong Kong, Shenzhen. G.~Y.~S.~was supported by the NSFC under the Grant No.
11704186 and the startup Fund of Nanjing University of Aeronautics and Astronautics under the Grant No.~YAH17053. M.~J.~H.~was supported by the ERC Synergy grant BioQ. Numerical simulations were carried out on the clusters at National Supercomputing Center in Shenzhen and Nanjing University of Aeronautics and Astronautics.
\end{acknowledgements}

\end{document}